\documentclass{article}
\usepackage{graphicx}
\usepackage{amsmath}

\begin{document}

\title{Partial Wave Analysis of the First Order Born Amplitude of a Dirac particle
in an Aharonov-Bohm Potential}
\author{M.S. Shikakhwa \\
Department of Physics,\\
University of Jordan, Amman-Jordan\\
and\\
N.K. Pak\\
Department of Physics,\\
Middle East Technical University,\\
06531 Ankara-Turkey.}
\maketitle

\begin{abstract}
A partial wave analysis using the basis of the total angular momentum
operator $J_{3}$ is carried out for the first order Born amplitude of a Dirac
particle in an Aharonov-Bohm (AB) potential. It is demonstrated that the
s-partial wave contributes to the scattering amplitude in contrast to the case
with scalar non-relativistic particles. We suggest that this explains the
fact that the first order Born amplitude of a Dirac particle coincides with
the exact amplitude expanded to the same order, where it does not for a
scalar particle. An interesting algebra involving the Dirac velocity operator and
the angular observables is discovered and its consequences are exploited.
\end{abstract}

\begin{center}
\bigskip
\end{center}

\section{Introduction}

The first attempts to calculate the Aharonov-Bohm (AB) scattering\cite{ab} amplitude for a
scalar particle using perturbation theory \cite{Feinberg,corinaldesi}
revealed a discrepancy between the first order Born amplitude and the exact
amplitude when expanded to the same order.Moreover, the second order Born
amplitude turned out to be divergent. These results were attributed \cite{corinaldesi}
to the fact that the first order Born amplitude based on the Schr\"{o}dinger
Hamiltonian of a scalar particle misses the contribution of the $l=0$
partial wave, as it is of second order. The problem manifested itself
also in the field theory models of the AB effect with scalar particles,
namely the Chern-Simons models\cite{bergman}. It also appeared in the perturbative
calculations in the many-body anyon theories near the bosonic end\cite{Ouvry}. It
was noted that introducing a contact interaction into the Hamiltonian
remedies these problems\cite{bergman}.
Subsequently this interaction was attributed
to a spin-magnetic moment interaction\cite{comtet}. The first order Born amplitude for
a Dirac particle was calculated in \cite{vera} and the second order in \cite{boz} where
full agreement with the expansions of the exact amplitude\cite{hagen1} to the
corresponding order was found. Non-relativistic perturbative calculations
within the framework of the field theory models of the AB with spin-1/2 particles
suffered no problems\cite{hagen2,vfain}.

No partial wave analysis of the first order Born amplitude for a Dirac
particle where it would be interesting to investigate the behavior of the
$l=0$ partial wave was reported in the literature. The main
motivation behind this work is to carry out such an analysis.

In section 2, we present a comprehensive partial wave analysis of the first
order Born amplitude for non-relativistic scalar and spin 1/2 particles. In
section 3, we carry out a partial wave analysis of the Born amplitude of a
Dirac particle, using the cylindrical partial modes of the conserved total
angular momentum operator. An interesting closed algebra involving the Dirac
velocity operator and the angular observables of the theory is discovered,
and its consequences pursued.

\section{Partial wave Born amplitude for non-relativistic scalar and spin-1/2 particles}

Before embarking on the treatment of the Dirac particle, we will first carry out
the partial wave analysis for the non-relativistic scalar and spin-1/2 particles in
the AB potential. While the results of this discussion are generally known
and were mentioned in the literature in various contexts\cite{comtet}, there is no published
work that we know of, which contains a systematic and complete treatment.
Thus we present it here for completeness and to set the stage for the discussion of
the relativistic case.

The AB potential in the cylindrical coordinates reads
\begin{equation}
\mathbf{A}=
\frac{\Phi }{2\pi \rho }\mathbf{\epsilon}_{\varphi},
\label{eq1}
\end{equation}
where $\rho =\sqrt{x^{2}+y^{2}}$,
$\mathbf{\epsilon}_{\varphi}$ is the unit vector along the $\varphi$
direction and $\Phi $ is the flux through the tube. The
Schr\"{o}dinger equation for a scalar particle in this potential, written in
cylindrical coordinates, is ($\hbar =c=1$):
\begin{equation}
\left[ \frac{\partial ^{2}}{\partial \rho ^{2}}+\frac{1}{\rho }\frac{%
\partial }{\partial \rho }+\frac{1}{\rho ^{2}}(\frac{\partial }{\partial
\varphi }+i\alpha )^{2}+\frac{\partial ^{2}}{\partial z^{2}}+k^{2}\right]
\Psi (\mathbf{r)=}0,
\label{eq2}
\end{equation}
where $\alpha = -\frac{e\Phi}{2\pi}$. We take \( 0 < \alpha < 1, \)
as in this work we will be mainly interested in perturbative
calculations.

As usual, one can separate the z-dependence of the wavefunction and neglect
it all together without any loss of generality. The interaction potential in Eq.~(\ref{eq2})
can be identified as
\begin{equation}
U=-\frac{1}{\rho ^{2}}(2i\alpha \frac{\partial }{\partial \varphi })+\frac{%
\alpha ^{2}}{\rho ^{2}}.
\end{equation}

The first order Born scattering amplitude can now be readily constructed, and
reads
\begin{equation}
f^{\left( 1\right) }(\theta )=\left( \frac{i}{2(2\pi ik)^{\frac{1}{2}}}%
\right) \int e^{-i\mathbf{k}^{\prime }.\mathbf{x}}\left( \frac{2i\alpha }{%
\rho ^{2}}\frac{\partial }{\partial \varphi }\right) e^{i\mathbf{k}.\mathbf{x%
}}\rho d\rho d\varphi,
\label{eq4}
\end{equation}
where $\mathbf{k}$ and $\mathbf{k}^{\prime }$ are, respectively, the wave
vectors of the incident (from left) and scattered waves, with $\left| \mathbf{k}\right|
=\left| \mathbf{k}^{\prime }\right| $; and $\theta $ is the scattering angle. A
calculation of $f^{\left( 1\right) }(\theta )$  yields \cite{Feinberg,corinaldesi}:
\begin{equation}
f^{\left( 1\right) }(\theta )=-\alpha \left( \frac{\pi }{2ik}\right) ^{\frac{1%
}{2}}\frac{\cos \frac{\theta }{2}}{\sin \frac{\theta }{2}},\qquad \theta
\neq 0.
\end{equation}
The exact amplitude  first calculated in \cite{ab}, and corrected in
\cite{hagen},  for \( 0 < \alpha < 1 \)  reads:
\begin{equation}
f(\theta) = - \frac{i}{\sqrt{2\pi ik}} (\sin \pi\alpha) \frac{e^{-i
\theta/2}}{\sin \frac{\theta}{2}}
\end{equation}

\noindent
For small $\alpha$, one gets,
\begin{equation}
f(\theta )=\left( \frac{\pi }{2ik}\right) ^{\frac{1}{2}}\left( -\alpha \cot
\frac{\theta}{2}-i\alpha \right) +O\left( \alpha ^{2}\right) ,
\qquad \theta \neq 0.
\label{eq7}
\end{equation}
$f^{\left( 1\right) }(\theta )$ given in Eq. (5) clearly misses the -$i\alpha$ term of Eq.~(\ref{eq7}).
This discrepancy was attributed to the fact that the first
order Born amplitude misses the contribution of the s-partial wave \cite{corinaldesi}.
This can be seen most transparently by looking at the partial Born
amplitudes separately, to which we will now turn.

The plane waves in Eq.~(\ref{eq4}) can be expanded in terms of the conserved orbital
angular momentum operator $L_{3}$ by employing the well-known expansion
\begin{equation}
e^{ikx\cos \alpha }=\sum_{l=-\infty }^{+\infty } i^{l}e^{il\alpha }J_{l}\left( x\right),
\end{equation}
where $J_{l}\left( x\right) $ are the Bessel functions of order $l$.
After carrying out the angular integral in Eq.~(\ref{eq4}), we get:
\begin{equation}
f^{\left( 1\right) }(\theta )=\left( \frac{i\alpha }{\left( 2\pi ik\right) ^{%
\frac{1}{2}}}\right) \sum_{l}le^{il\theta }\int \frac{d\rho }{\rho }\left[
J_{l}\left( k\rho \right) \right] ^{2}
\end{equation}

Now, it is obvious that the $l=0$ partial wave amplitude, i.e., $f_{0}^{\left(
1\right) }(\theta )$ vanishes. Integrating over the Bessel functions with
the aid of the formula
\begin{equation}
\int_{0}^{\infty }dr\frac{\left[ J_{l}\left( r\right) \right] ^{2}}{r}%
=\left| 2l\right| ^{-1},\qquad l\neq 0,
\end{equation}
we get
\begin{equation}
f^{\left( 1\right) }(\theta )= -\sum_{l} {}^{\prime}i\alpha \pi
 \left( \frac{1}{2\pi ik}\right) ^{\frac{1}{2}} \mathrm{sgn}(l) e^{il\theta },
\end{equation}
where $\mathrm{sgn}\left(l\right)=\frac{l}{\left| l\right| }$, and the prime
denotes that the $l=0$ term is excluded from the summation. Recalling that
generally
\begin{equation*}
f^{\left( 1\right) }(\theta )=\sum_{l}^{{}}f_{l}^{\left( 1\right) }(\theta
)e^{il\theta },
\end{equation*}
we get the partial amplitudes as:
\begin{equation}
f_{l}^{\left( 1\right) }\left( \theta \right) =\left\{
\begin{array}{cl}
\frac{-i\pi \alpha }{\left( 2\pi ik\right) ^{\frac{1}{2}}} \mathrm{sgn}(l) \qquad ,\qquad
&l \neq 0 \\
0\qquad ,\qquad &l=0.
\end{array}
\right.
\label{eq12}
\end{equation}

To compare the above partial amplitudes with the exact ones expanded
in terms of \( \alpha \), we note that the exact phase shifts reported in
\cite{ruij, hagen} read (when \( 0 < \alpha < 1), \)

$$
\delta_{m} = \left\{\begin{array}{lll}
-\frac{\pi}{2} \alpha & , & m \ge 0\\ [.7em]
\frac{\pi}{2} \alpha & , & m < 0
\end{array}\right.
$$

Therefore, the exact partial amplitudes become  \cite{ruij}:
\begin{equation}
f_{l}\left( \theta \right) =\left( e^{-i\mathrm{sgn}\left(
l\right) \pi \alpha }-1\right) \left( 2\pi ik\right) ^{-\frac{1}{2}}\qquad
,\qquad l=0,\pm 1,\pm 2,....
\end{equation}
which, for small $\alpha$ reduce to Eq.~(\ref{eq12}) when $l\neq 0$.
When $l=0$, $f_{0}^{{}}\left( \theta \right) $ reduces to
 $\frac{-i\alpha\pi}{\left( 2\pi ik\right)^{\frac{1}{2}}}$ for small $\alpha$,
while $f_{0}^{\left( 1\right) }\left( \theta \right) $ vanishes!

We turn now to the non-relativistic spin-1/2 particles, where we will
see that the $l=0$ partial amplitude is non-vanishing.  In addition to
this, it will turn out that, it is this partial amplitude that
leads to the modification of the exact amplitude when the spin is
included.

The starting point is the Pauli equation
\begin{equation}
\frac{1}{2m}\left( \boldsymbol{\sigma} \cdot \boldsymbol{\Pi} \right)
^{2}\Psi=E\Psi ,
\end{equation}
where $\boldsymbol{\Pi} =\left( \mathbf{p-}e\mathbf{A}\right)$, and $\mathbf{A}$
is the AB potential given in Eq.~(\ref{eq1}), and \( \sigma^i , \ i =
1,2,3, \) are the Pauli spin matrices. Suppressing again the $z$ degree of
freedom we get,
\begin{equation}
\left[ \frac{\partial ^{2}}{\partial \rho ^{2}}+\frac{1}{\rho }\frac{%
\partial }{\partial \rho }+\frac{1}{\rho ^{2}}(\frac{\partial }{\partial
\varphi }+i\alpha )^{2}-2\pi \alpha \sigma _{3}\delta \left( \mathbf{r}%
\right) +k^{2}\right] \Psi (\mathbf{r)=}0\ .
\end{equation}

The first order Born amplitude now reads,
\begin{equation}
f^{\left( 1\right) }(\theta )=\left( \frac{i}{2(2\pi ik)^{\frac{1}{2}}}%
\right) \int e^{-i\mathbf{k}^{\prime }.\mathbf{x}}\chi ^{\dagger \left(
s^{\prime }\right) }\left( \frac{2i\alpha }{\rho ^{2}}\frac{\partial }{%
\partial \varphi }-2\pi \alpha \sigma _{3}\delta \left( \mathbf{r}\right)
\right) \chi ^{\left( s\right) }e^{i\mathbf{k}.\mathbf{x}}\rho d\rho
d\varphi ,
\end{equation}
where
$\chi^{\left(s\right)}$ and $\chi ^{\left(s^{\prime }\right)}$ are the
spinors of the incident and outgoing waves, respectively. Expanding the
plane waves, and carrying out the
integrals as before, we get:
\begin{equation}
f^{\left( 1\right) }(\theta )=\frac{1}{\left(2\pi ik\right) ^{\frac{1}{2}}}
\sum_{l}^{{}}e^{il\theta }\chi^{\dagger\left( s^{\prime }\right) }
\left[  -i\pi \alpha \mathrm{sgn}\left(l\right)
\left( 1-\delta _{l,0}\right) -i\pi \alpha \delta _{l,0}\sigma _{3} \right]
\chi ^{\left( s\right) }.
\end{equation}

Taking $\chi ^{\left( s\right) }$ to be the spin state of a particle
polarized along an arbitrary direction specified by a unit vector $\mathbf{n}$
with polar angle $\beta$,
and considering transitions to a final state polarized along the same
direction, we get the amplitude as:
\begin{equation}
f^{\left( 1\right) }(\theta )=\sum_{l}
\left( 2\pi ik\right) ^{-\frac{1}{2}}
e^{il\theta }\left[ -i\pi
\alpha \mathrm{sgn}\left( l\right) \left( 1-\delta _{l,0}\right) -i\pi \alpha \delta
_{l,0}\cos\beta
\right],
\end{equation}
\begin{equation}
f_{l}^{\left( 1\right) }\left( \theta \right) =\left\{
\begin{array}{c}
\frac{-i\pi \alpha }{\left( 2\pi ik\right) ^{\frac{1}{2}}}\mathrm{sgn}(l) \qquad ,\qquad l \neq 0 \\
-\frac{i\pi \alpha }{\left( 2\pi ik\right) ^{\frac{1}{2}}}\cos\beta \qquad
,\qquad l=0\ .
\end{array}
\right.
\end{equation}

The above results demonstrate that the $l=0$ partial amplitude is
non-vanishing, the reason being the spin-magnetic moment interaction term.
We also note that for our choice of the spin orientations, it is only the
s-wave that flips the spin, modifying the unpolarized amplitude only when the incident
particle's spin has a component perpendicular to the solenoid. This result
was first reported in \cite{hagen1} for the exact amplitude, and verified for the
first-order Born amplitude in \cite{vera}. This is quite natural, as the s-wave is
the only partial wave that can feel the solenoid; the other waves being
banned by the centrifugal barrier.

\section{Partial Wave Born Amplitudes for a Dirac Particle}

The Hamiltonian for a Dirac particle in an electromagnetic potential is:
\begin{equation}
H=H_{\circ}+H_{int} ,
\end{equation}
where
\begin{equation}
H_{\circ}=\boldsymbol{\alpha}\cdot\mathbf{p}+\beta m,
\end{equation}
and
\begin{equation}
H_{int}=eA_{\circ} - e\boldsymbol{\alpha}.\mathbf{A}
\end{equation}
Here, $\alpha _{i}=\beta \gamma _{i}$ and $\beta =\gamma _{4}$.
The $\gamma$'s are the Dirac matrices: $\left\{ \gamma _{\mu },\gamma _{\nu
}\right\} =2g_{\mu \nu }$.

The first-order Born amplitude for the scattering of a Dirac particle in an
electromagnetic field then reads,
\begin{equation}
S_{fi}^{\left( 1\right) }=-i \int d^{4}x\overline{\psi}
_{f}^{\left( s^{\prime }\right) }(\vec{x})\left(e\gamma _{\mu }A^{\mu }\right)
\psi_{i}^{\left( s\right) }(x).
\end{equation}
With the AB potential as given in Eq.~(\ref{eq1}), and with the choice of
gauge $A_0=0$, and suppressing the $z$ degree of freedom, and an
energy consering \( \delta \)-function, we get
\begin{equation}
S_{fi}^{\left( 1\right) }=i\alpha \int d\rho d\varphi
\overline{\psi}_{f}^{\left( s^{\prime }\right) }(\vec{x})\left( -\sin \varphi \gamma
_{1}+\cos \varphi \gamma _{2}\right) \psi_{i}^{\left( s\right) }(\vec{x}).
\label{eq24}
\end{equation}
where $p^{\bot}$ is the magnitude of the momentum perpendicular to the solenoid.
For later convenience, we write $S_{fi}^{\left( 1\right) }$ as :
\begin{equation}
S_{fi}^{\left( 1\right) }=i\alpha \int d\rho d\varphi
\overline{\psi}_{f}^{\left( s^{\prime }\right) }(\vec{x})
\left( D^{+}+D^{-}\right)
\psi_{i}^{\left( s\right) }(\vec{x}). 
\end{equation}
where the operators D$^{\pm }$ are defined by:
\begin{equation}
D^{\pm }=\left( \frac{\gamma _{2}\pm i\gamma _{1}}{2}\right) e^{\pm i\varphi}.
\end{equation}

Prior to carrying out a partial wave analysis of (\ref{eq24}), we
have to note that an expansion of the incident and outgoing waves in terms
of the $L_{3}$ eigenstates will be inconclusive in this case. The reason, physically
speaking, is that $L_{3}$ is not a constant of the motion in the Dirac theory,
not even (as is well-known) in the free theory. The spinors, $u_{i}^{\left(
s\right) }$ and $u_{f}^{\left( s\right) }$ are now functions of the angle $%
\varphi$. So, one has to expand the free spinors in terms of the eigenstates
of the conserved total angular momentum operator $J_{3}=L_{3}+\frac{\Sigma
_{3}}{2}$. We need first to find these states. These will be taken to be
simultaneous eigenstates of the set of commuting operators: $H_{\circ
},J_{3},S_{3}=\beta \Sigma _{3}+\frac{\xi p_{3}}{m}$ and $p_{3}$ (where
$\xi =\left(
\begin{array}{cc}
0 & I \\
I & 0
\end{array}
\right)$) according to:
\begin{equation}
\begin{array}{cc}
H_{\circ }\Psi _{ls}= & E\Psi _{ls} \\
J_{3}\Psi _{ls}= & \left( l+\frac{1}{2}\right) \Psi _{ls} \\
p_{3}\Psi _{ls}= & p_{3}\Psi _{ls} \\
S_{3}\Psi _{ls}= & \pm s\Psi _{ls}
\end{array}
\label{eq27}
\end{equation}

Here, we are diagonalizing the spin operator $S_{3}$ along with the
Hamiltonian rather than the more conventional helicity operator. $S_{3}$ is
usually used when one has a magnetic field along the $z$-axis\cite{sokolow}.
In the non-relativistic limit the upper components of $\Psi_{ls} $
are eigenstates of $\sigma _{3}$. The eigenvalues of $S_{3}$ are
\begin{equation}
s=\pm \sqrt{1+\left( \frac{p_{3}}{m}\right) ^{2}},
\label{eq28}
\end{equation}
which reduce to $s=\pm 1$ when $p_{3}$ is set to zero. The $\Psi _{ls}$ that
solve the set of equations (\ref{eq27}) read
\begin{equation}
\Psi _{ls}=\frac{e^{-i(Et-p_{3}x_{3}-l\varphi )}}{\sqrt{2\pi} \sqrt{2E}\sqrt{2s}}%
\left(
\begin{array}{c}
\sqrt{E+sm}\sqrt{s+1}J_{l}\left( p^{\bot }\rho \right) \\
ie^{i\varphi }\epsilon _{3}\sqrt{E-sm}\sqrt{s-1}J_{l+1}\left( p^{\bot }\rho
\right) \\
\epsilon _{3}\sqrt{E+sm}\sqrt{s-1}J_{l}\left( p^{\bot }\rho \right) \\
ie^{i\varphi }\sqrt{E-sm}\sqrt{s+1}J_{l+1}\left( p^{\bot }\rho \right)
\end{array}
\right) ,
\end{equation}
where $\epsilon_{3}=\mathrm{sgn}(s)\mathrm{sgn}(p_{3})$,$p^{\bot}$
is the magnitude of the momentum perpendicular to the solenoidand,
and $s$ assumes the values given in Eq.~(\ref{eq28}). Setting
$p_{3}$ to zero one gets $\Psi _{ls}$ modes as:
\begin{equation}
\Psi _{ls}(\mathbf{x})=\frac{e^{i(l\varphi )}}{\sqrt{2\pi }\sqrt{2E}\sqrt{2s}%
}\left(
\begin{array}{c}
\sqrt{E+sm}\sqrt{s+1}J_{l}\left( p^{\bot }\rho \right) \\
ie^{i\varphi }\epsilon _{3}\sqrt{E-sm}\sqrt{s-1}J_{l+1}\left( p^{\bot }\rho
\right) \\
\epsilon _{3}\sqrt{E+sm}\sqrt{s-1}J_{l}\left( p^{\bot }\rho \right) \\
ie^{i\varphi }\sqrt{E-sm}\sqrt{s+1}J_{l+1}\left( p^{\bot }\rho \right)
\end{array}
\right) ,
\label{eq30}
\end{equation}
where $s=\pm 1$ now, and the cylindrical partial modes $\Psi _{ls}(\mathbf{x})$
are normalized as

\begin{equation}
\int\rho d\rho d\varphi \Psi _{l^{\prime }s^{\prime }}^{\dagger }(\mathbf{x}%
)\Psi _{ls}(\mathbf{x})=\frac{1}{p^{\bot }}\delta \left( p^{\bot }-p^{\bot
\prime }\right) \delta _{l,l^{\prime }}\delta _{s.s^{\prime }}.
\end{equation}

The above partial modes are now the correct expansion basis that
are to be used in the partial wave analysis. The incident and
outgoing waves which are also  eigenstates  of $S_{3}$ are\((p_{3}
= 0, \ s = \mp 1): \)
\begin{equation}
\Psi ^{(s)}_{i}(\mathbf{x})=e^{i\mathbf{p}_{i}.\mathbf{x}}u_{i}=\frac{e^{ip^{\bot
}\rho \cos \varphi }}{\sqrt{4\pi}\sqrt{2s}}\left(
\begin{array}{c}
\sqrt{E+sm}\sqrt{s+1} \\
\epsilon _{3}\sqrt{E-sm}\sqrt{s-1} \\
\epsilon _{3}\sqrt{E+sm}\sqrt{s-1} \\
\sqrt{E-sm}\sqrt{s+1}
\end{array}
\right) ,
\end{equation}
\begin{equation}
\Psi _{f}^{\left( s\right) }(\mathbf{x})=e^{i\mathbf{p}_{f}.\mathbf{x}}u_{f}=%
\frac{e^{ip^{\bot }\rho \cos \left( \varphi -\theta \right)
}}{\sqrt{4\pi}\sqrt{2s}}\left(
\begin{array}{c}
\sqrt{E+sm}\sqrt{s+1} \\
\epsilon _{3}e^{i\theta }\sqrt{E-sm}\sqrt{s-1} \\
\epsilon _{3}\sqrt{E+sm}\sqrt{s-1} \\
e^{i\theta }\sqrt{E-sm}\sqrt{s+1}
\end{array}
\right) ,
\end{equation}
The incident and outgoing waves given in (32) and (33) are
normalized as \( \int d^2 x \psi^{\dagger(s')}{(\vec{x})}
\psi^{(s)}{(\vec{x})}= E \delta (\vec{p} - \vec {p'}) \), which is
the Lorentz-invariant normalization.

We can verify the following expansion of $\Psi _{i}(\mathbf{x})$ and $\Psi
_{f}(\mathbf{x})$ in terms of the cylindrical modes $\Psi _{ls}(\mathbf{x}):$
\begin{equation}
\begin{array}{ll}
\Psi _{i}^{\left( s\right) }(\mathbf{x})= & \sqrt{E_{i}} \sum_{l}\left( i\right) ^{l}\Psi
_{ls}(\mathbf{x}) \\
\Psi _{f}^{\left( s\right) }(\mathbf{x})= & \sqrt{E_{f}} \sum_{l}\left( i\right)
^{l}e^{-il\theta }\Psi _{ls}(\mathbf{x})\ .
\end{array}
\end{equation}

The amplitude $S_{fi}^{\left( 1\right) }$ now takes the form
\begin{equation}
S_{fi}^{\left( 1\right) }=i\alpha E \sum_{l}\left( i\right)
^{l}\sum_{l^{\prime }}\left( -i\right) ^{l^{\prime }}e^{il^{\prime }\theta }%
\mathcal{M},
\end{equation}
where
\begin{equation}
\mathcal{M=}\int d\rho d\varphi \overline{\Psi }_{l^{\prime }s^{\prime }}(%
\mathbf{x})\left( D^{+}+D^{-}\right) \Psi _{ls}(\mathbf{x}),
\label{eq36}
\end{equation}
and \( E = E_{i} = E_{f} \).

Now, the operators $D^{\pm}$, being linear combinations of the
$\gamma$ matrices will flip the spinors $\Psi _{ls}$. On the other hand,
since $\left[ J_{3},D^{\pm }\right] =\left[ S_{3},D^{\pm }\right] =0$, then $%
D^{\pm }\Psi _{ls}$ should still be eigenstates of $J_{3}$ and $S_{3}$. It
turns out that $D^{\pm }$ operators together with the angular observables of the theory
obey an interesting algebra which leads to a fulfillment of the the
above requirements. Let us first note that $D^{\pm }\Psi _{ls}$ are
eigenstates of $L_{3}$ and $\frac{\Sigma _{3}}{2}$ as can be verified
directly, though $\Psi _{ls}$ obviously is not
\begin{equation}
\begin{array}{cc}
\frac{\Sigma _{3}}{2}D^{\pm }\Psi _{ls}= & \mp \frac{1}{2}D^{\pm }\Psi _{ls} \\
L_{3}D^{\pm }\Psi _{ls}= & \left(
\begin{array}{c}
l+1 \\
l
\end{array}
\right) D^{\pm }\Psi _{ls}\ .
\end{array}
\label{eq37}
\end{equation}
It follows from Eq.~(\ref{eq37})that
\begin{equation}
\left( L_{3}+\frac{\Sigma _{3}}{2}\right) D^{\pm }\Psi _{ls}=\left(
\begin{array}{c}
\left( -\frac{1}{2}\right) +\left( l+1\right) \\
\left( +\frac{1}{2}\right) +\left( l\right)
\end{array}
\right) D^{\pm }\Psi _{ls}=\left( l+\frac{1}{2}\right) D^{\pm }\Psi _{ls}
\end{equation}
as should be. Therefore, the operators $D^{\pm }$ acting on the $\Psi _{ls}$
modes, project them into eigenstates of $L_{3}$ and $\frac{\Sigma _{3}}{2}$
such that the sum of the eigenvalues is always equal to the $J_{3}$
eigenvalue; $l+\frac{1}{2}$. To get a further insight into the mechanism in
action, we first note that the $\Psi _{ls}$ modes can be written as linear
combinations of the eigenstates of \ the $L_{3}$ and $\frac{\Sigma _{3}}{2}$
operators. Explicitly:
\begin{equation}
\Psi _{ls=+1}=\frac{1}{\sqrt{4\pi E}}\left[\left(
\begin{array}{c}
\sqrt{E+m}J_{l}\left( p^{\bot }\rho \right) e^{i(l\varphi )} \\
0 \\
0 \\
0
\end{array}
\right) +\left(
\begin{array}{c}
0 \\
0 \\
0 \\
i\sqrt{E-m}J_{l+1}\left( p^{\bot }\rho \right) e^{i(l+1)\varphi }
\end{array}
\right)\right],
\end{equation}
or in a more compact notation
\begin{equation}
\left| j_{3},s=1\right\rangle =\left| j_{3},s=1;\,l,+\frac{1}{2}%
\right\rangle +\left| j_{3},s=1;\,l+1,-\frac{1}{2}\right\rangle ,
\label{eq40}
\end{equation}
where, the quantum numbers ($l,l+1$) and ($\frac{1}{2},-\frac{1}{2}$) above refer to
the eigenvalues of \ $L_{3}$ and $\frac{\Sigma _{3}}{2}$, respectively, and
the total orbital angular momentum's quantum number is always $j_{3}$=$l+%
\frac{1}{2}$. Similarly, for $s=-1$, we have
\begin{equation}
\left| j_{3},s=-1;\right\rangle =\left| j_{3},s=-1;\,l+1,-\frac{1}{2}%
\right\rangle +\left| j_{3},s=-1;\,l,+\frac{1}{2}\right\rangle .
\label{eq41}
\end{equation}

One can verify the following algebra
\begin{equation}
\begin{array}{cc}
\left[ L_{3},D^{\pm }\right] = & \pm D^{\pm } \\
\left[ \frac{\Sigma _{3}}{2},D^{\pm }\right] = & \mp D^{\pm } \\
\left[ D^{+},D^{-}\right] = & 2\left( \frac{\Sigma _{3}}{2}\right)\ .
\end{array}
\end{equation}
Note also that
\begin{equation}
\left( D^{+}\right) ^{2}=\left( D^{-}\right) ^{2}=0.
\label{eq43}
\end{equation}

This algebra means that the operators $D^{\pm }$ are some sort of raising
and lowering operators in the angular momentum space of the theory. Indeed,
denoting the simultaneous eigenstate of
$L_{3}$ and $\frac{\Sigma _{3}}{2}$ as $\left| l_{3},\sigma _{3}\right\rangle $, one has:
\begin{equation}
\begin{array}{cc}
L_{3}D^{\pm }\left| l_{3},\sigma _{3}\right\rangle = & \left( l_{3}\pm
1\right) D^{\pm }\left| l_{3},\sigma _{3}\right\rangle \\
\frac{\Sigma _{3}}{2}D^{\pm }\left| l_{3},\sigma _{3}\right\rangle = & \left(
\sigma _{3}\mp 1\right) D^{\pm }\left| l_{3},\sigma _{3}\right\rangle .
\end{array}
\end{equation}
Therefore, $D^{\pm }\left| l_{3},\sigma _{3}\right\rangle =c_{\pm }\left|
\left( l_{3}\pm 1\right) ,\left( \sigma _{3}\mp 1\right) \right\rangle $.
The complex numbers $c_{\pm }$ are readily verified to be pure phases which
we set to 1. Moreover, note that Eq.~(\ref{eq43}) implies
\begin{equation}
D^{+}\left| l_{3},\sigma _{3}=-\frac{1}{2}\right\rangle =D^{-}\left|
l_{3},\sigma _{3}=+\frac{1}{2}\right\rangle =0.
\label{eq45}
\end{equation}
Thus, we have
\begin{equation}
D^{\pm }\left| l_{3},\sigma _{3}\right\rangle =\left| l_{3}\pm 1,\sigma
_{3}\mp 1\right\rangle .
\label{eq46}
\end{equation}

Going back to our $\Psi _{ls}$
functions given in Eqs.~(\ref{eq40}) and (\ref{eq41}), we see now that
\begin{equation}
\left(
\begin{array}{c}
D^{+} \\
D^{-}
\end{array}
\right) \left| j_{3},s=1\right\rangle =\left(
\begin{array}{c}
\left| j_{3},s=1;\,l+1,-\frac{1}{2}\right\rangle \\
\left| j_{3},s=1;\,l,+\frac{1}{2}\right\rangle
\end{array}
\right) ,
\end{equation}
and
\begin{equation}
\left(
\begin{array}{c}
D^{+} \\
D^{-}
\end{array}
\right) \left| j_{3},s=-1\right\rangle =\left(
\begin{array}{c}
\left| j_{3},s=-1;\,l+1,-\frac{1}{2}\right\rangle \\
\left| j_{3},s=-1;\,l,+\frac{1}{2}\right\rangle
\end{array}
\right) .
\end{equation}

This means that the operators $D^{\pm }$ acting on $\left| j_{3},s=\pm
1\right\rangle $ projects out eigenstates of $L_{3}$ and $\frac{\Sigma _{3}}{%
2}$ such that $l_{3}+\sigma _{3}=l+\frac{1}{2}$ only, i.e. $J_{3}$
eigenstates.

This mechanism of conserving the $J_{3}$ quantum number can be only observed
upon employing the partial wave expansion of the Dirac spinors.

Going back to our amplitude; upon substituting the explicit forms
of the partial modes of Eq.~(\ref{eq30}), in Eq.~(\ref{eq36}), and
carrying out the $\varphi $ integral we finally get,

\begin{equation}
\mathcal{M=}\left( \pi \right)
\frac{\sqrt{E^{2}-s^{2}m^{2}}
}{2E}
\left( 2s\right)
\delta _{l,l^{\prime }}\delta _{s,s^{\prime }}\int J_{l+1}\left( p^{\bot
}\rho \right) \,J_{l}\left( p^{\bot }\rho \right) d\rho  .
\end{equation}

The above expression clearly conserves $J_{3}$ and $S_{3}$ quantum numbers
as it should do. Moreover, the $l=0$ partial wave contributes to the
amplitude on equal footing with the other partial waves.

The Bessel function integral in Eq. (49) is tabulated for positive
values of \( \ell \) (formula 6.512-3 in \cite{grads}). For negative
values of \( \ell \), we make use of the well-known relation valid for
integral \( \ell \), \( J_{-\ell}(x) = (-1)^{\ell} J_{\ell}(x) \), so
that we convert the integral over Besssel functions of negative
order to an integral over Bessel functions of positive order, getting
an overall minus sign. So, we get finally for the first order
amplitude,

\begin{equation}
S_{fi}^{(1)} = \sum_{\ell} \frac{1}{2} i \alpha sgn(\ell)
e^{i\ell\theta}
\end{equation}
The partial amplitudes are therefore,

\begin{equation}
S_{\ell}^{(1)} = \frac{1}{2}i\alpha sgn(\ell) , \ \ \ell = 0 , \mp 1,
\mp 2,\ldots
\end{equation}
Note the appearance of sgn(\( \ell \)) which resulted from the
Bessel function integral in Eq. (49). This is same as in the
non-relativistic amplitude. To compare our final expression in Eq.
(51) with the non-relativistic partial scattering amplitudes, \(
f_{\ell}^{(1)} \) \( (\theta) \), we note that the S-matrix and
the scattering amplitude are related in two dimensions via
\cite{ruij}:

\begin{equation}
(S-1) (k, \theta) = \left(\frac{ik}{2\pi}\right)^{1/2} f(k, \theta).
\end{equation}

Expanding \( S(k, \theta) \) and \( f(k, \theta) \) in powers of
the coupling constant, and imposing the equality for each partial
wave, we get

\begin{equation}
f_{\ell}^{(1)} (k, \theta) = \sqrt{\frac{2\pi}{ik}} S_{\ell}^{(1)}
(k, \theta)
\end{equation}
Substituting \( S_{\ell}^{(1)} \) given in Eq. (51), in Eq. (53), we
get the partial scattering amplitudes:

\begin{equation}
f_{\ell}^{(1)} (\theta) = (2\pi ik)^{-1/2} i\alpha\pi sgn
(\ell), \ \ 1 = 0, \mp 1, \mp 2,\ldots
\end{equation}

Eq. (54) compares (up to an overall minus sign) with Eq. (19). The
discrepancy for the partial amplitudes \( f_{0}^{(1)} (\theta) \) is a
result  of the difference in the spin orientations of the incident
and outgoing particles in the two cases.


\section{Conclusions}

We have demonstrated through an explicit partial wave analysis, that the
inclusion of spin into the Hamiltonian of a non-relativistic particle in an
AB field leads to a non-vanishing $l=0$ first order partial Born amplitude.
Moreover, this particular amplitude is the one responsible for the
modification of the total amplitude reported in \cite{hagen1} as a result of the
inclusion of spin. A partial wave analysis of the first order Born amplitude
for a Dirac particle shows that all the partial amplitudes, including the
$l=0$ are non-vanishing and contribute equally to the total amplitude. An
interesting algebra involving the Dirac velocity operator and the angular
observables of the Dirac theory was discovered, and shown to lead to a
mechanism for the conservation of the total angular momentum quantum number
upon transitions from the initial to the final states at the level of each
partial wave.

\section*{Acknowledgment}

M.S.S. thanks the Physics Department in the Middle East Technical
University, where this work has been carried out for hospitality,
and the The Scientific and Technical
Research Council of Turkey (T\"{U}B\.{I}TAK) for partial financial support.

\end{document}